\documentclass[10pt,a4paper,twoside]{article}
\usepackage{epsfig}
\usepackage{baltlat6}
\usepackage{array}
\usepackage{here}
\usepackage{hyperref}   
\usepackage{color}
\usepackage{amsmath,amssymb}
\def\aplt{\ {\raise-.5ex\hbox{$\buildrel<\over\sim$}}\ }

\begin{document}
\ \ \vspace{0.5mm} \setcounter{page}{367}

\titlehead{Baltic Astronomy, vol.\,24, 367--378, 2015}

\titleb{STATISTICAL ANALYSIS OF A COMPREHENSIVE LIST OF\\ VISUAL BINARIES}

\begin{authorl}
\authorb{D. Kovaleva,}{1}
\authorb{O. Malkov,}{1,2}
\authorb{L. Yungelson,}{1}
\authorb{D. Chulkov}{1} and
\authorb{G. M. Yikdem}{3}
\end{authorl}

\begin{addressl}
\addressb{1}{Institute of Astronomy, Russian Academy of Sciences,\\
48 Pyatnitskaya Str., 119017 Moscow, Russia;
dana@inasan.ru}
\addressb{2}{Faculty of Physics, M.\,V. Lomonosov Moscow State
University,\\
Leninskie Gory 1, Bld. 2, Moscow 119991, Russia}
\addressb{3}{Astronomy and Astrophysics Research Division,
Entoto Observatory and\\ Research Center, P.O. Box 33679 Addis
Ababa, Ethiopia}
\end{addressl}

\submitb{Received: 2015 November 2; accepted: 2015 November 30}

\begin{summary}
Visual binary stars are the most abundant class of observed
binaries. The most comprehensive list of data on visual binaries
compiled recently by cross-matching the largest catalogues of visual
binaries allowed a statistical investigation of observational
parameters of these systems. The dataset was cleaned by correcting
uncertainties and misclassifications, and supplemented with
available parallax data. The refined dataset is free from technical
biases and contains 3676 presumably physical visual pairs of
luminosity class V with known angular separations, magnitudes of the
components, spectral types, and parallaxes. We also compiled a
restricted sample of 998 pairs free from observational biases due to
the probability of binary discovery. Certain distributions of
observational and physical parameters of stars of our dataset are
discussed.
\end{summary}


\resthead{Statistical analysis of the set of visual binaries}
{D.~Kovaleva, O.~Malkov, L.~Yungelson, D.~Chulkov, G.\,M.~Yikdem}

\sectionb{1}{INTRODUCTION}

Visual binary stars are very important because they are the most
abundant observational type of binary systems.  The number of known
catalogued visual pairs (those that can be visually resolved using a
telescope) exceeds 130\,000. However, this dataset is not promising
enough for researchers because of quite limited amount of data
available per pair; typically, these include only the angular
separation between the components, $\rho$, and the positional angle,
$\theta$. However, the large number of such stars was believed to
justify the investigation of some properties of the entire
population of wide binaries based on these data. An exhaustive
analysis based on the data available by mid-1980's was performed by
Vereschagin et al. (1988). It involved the data for about 70\,000
visual pairs, which were used to derive the distributions of the
primary component mass, mass ratio, and semimajor axis of the orbit.

Recently, a new comprehensive set of data on visual binaries was
compiled by Isaeva et al. (2015) by cross-matching the current
version of The Washington Visual Double Star Catalog (WDS, Mason et
al. 2014), the Catalog of Components of Double \& Multiple stars
(CCDM, Dommanget \& Nys 2002), and the Tycho Double Star Catalogue
(TDSC, Fabricius et al. 2002). This list is named WCT, after the
first letters of abbreviations of the three source catalogues. Note
that the cross-matching of the TDSC catalogue with the WDS catalogue
is complete (i.e., all TDSC stars are included in the WDS), while no
match for 1872 pairs from the CCDM could be found in the WDS
catalogue. Thus, the primary WCT list contains almost 131\,000 pairs
and must grow in the future due to the systematic growth of the WDS.
We use for further analysis the observational data on positional
angle, angular separation, magnitudes, and spectra (when available)
of the components from all the three catalogues, preferring, where
possible, the data from the WDS if they are consistent with those
from the other sources, and make additional checks on what data to
choose in the case of any doubt or contradiction. Additionally, the
WCT contains parallaxes for more than 14\,000 pairs, mostly from the
Hipparcos catalogue (with the remaining ones adopted from SIMBAD).
This gives promise that we will be able not just to repeat the study
of Vereschagin et al. (1988), thereby based on a larger number of
objects and up-to-date data, but also to obtain high-quality
results.

In Section 2 we discuss the cleaning of the WST data from the errors
of the original catalogues. In Section 3 we describe the selection
of the data for statistical investigation. In Section 4 we present
some preliminary results of our investigation and briefly discuss
its prospects. Section 5 summarizes the conclusions.

\sectionb{2}{ERRONEOUS AND REDUNDANT DATA IN WCT SOURCE\\ CATALOGUES}
\vskip 1 mm

The WCT source catalogues are not error free. This especially
concerns the large compiled WDS and CCDM  catalogues. It is
desirable to fix this problem before whatever further statistical
analyses are performed to avoid eventual biases. Some types of
errors can be discovered without invoking external data sources.
These are, in particular, the cases where (i) discrepant positional
information is provided for an additional component, and (ii) a
binary or a component is listed twice, under different names.

\subsectionb{2.1}{Identifying erroneous data}

The catalogue entries usually provide the coordinates of the
reference ($\alpha_1, \delta_1$) and additional components
($\alpha_2, \delta_2$). In addition, the catalogues considered also
provide the position angle ($\theta$) and separation ($\rho$) for
the additional component in a pair, i.e., its position relative to
the reference component. From these data, the coordinates of the
additional component can be calculated as
\begin{equation}
\begin{aligned}
\alpha_3 - \alpha_1 &= \frac {\rho \sin \theta}{\cos \delta_1}, \\
\delta_3 - \delta_1 &= \rho \cos \theta \label{equ:additional}
\end{aligned}
\end{equation}
and compared to ($\alpha_2, \delta_2$). The angular separation
between two points in the sky can be calculated by the following
formula:
\begin{equation}
d = \sqrt {\cos^2 \delta_2 (\alpha_3 - \alpha_2)^2 + (\delta_3 - \delta_2)^2}.
\label{equ:distance}
\end{equation}
If the separation $d$ exceeds a certain adopted threshold $d_1$, the
data combination ($\alpha_1, \delta_1, \theta, \rho, \alpha_2,
\delta_2$) is considered to be inconsistent. The cutoff separation
$d_1$ depends on observational conditions, and we should estimate it
experimentally for every particular catalogue.


\begin{figure}[!tH]
\vbox{
\centerline{\psfig{figure=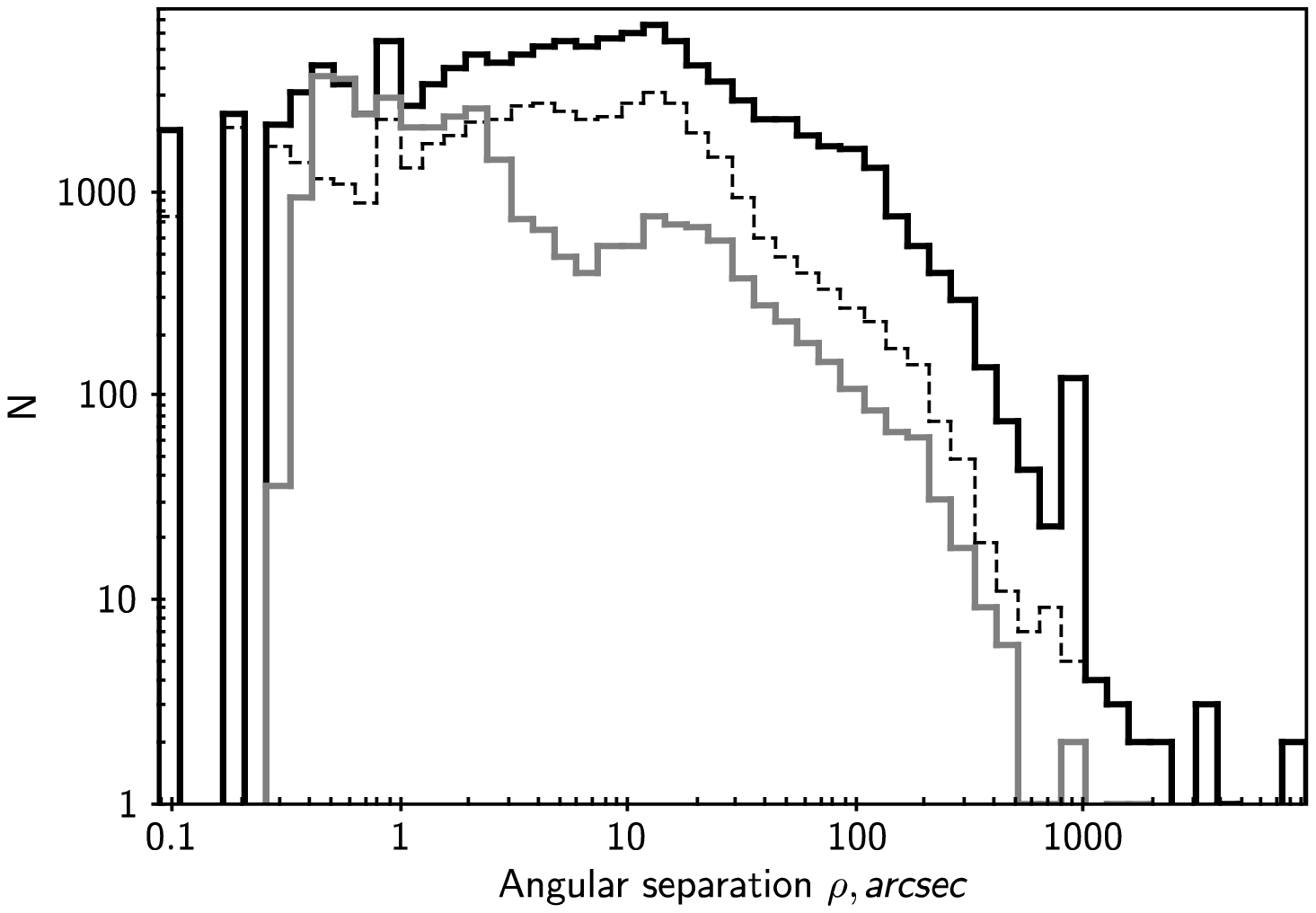,width=88mm,angle=0,clip=}}}
\vspace{1mm} \captionb{1} {Distribution of angular separations
$\rho$ of the WCT pairs. The solid, dashed, and gray lines show the
distributions for the WDS, CCDM, and TDSC data, respectively. The
peaks at 0.1, 1.0, and 1000~arcsec (for the WDS data) are false and
due to round-off procedures. }
\end{figure}

\subsectionb{2.2}{Fixing redundant data}

Some stars may be members of two different binary (multiple) systems
included in a compiled catalogue of binaries. Moreover, a binary can
be discovered by two different observers independently and,
consequently, this pair appears twice in the catalogue, under
different designations. This can be checked by coordinate
comparison. Similarly to the case described in \S\,2.1, the angular
separation between two points with celestial coordinates ($\alpha_x,
\delta_x$) and ($\alpha_y, \delta_y$) can be calculated as
\begin{equation}
d = \sqrt {\cos^2 \delta_x (\alpha_y - \alpha_x)^2 + (\delta_y - \delta_x)^2}
\label{equ:distance}
\end{equation}
and compared to a certain limiting value $d_2$.
If $d$ does not
exceed a certain threshold $d_2$, one can conclude that ($\alpha_x, \delta_x$)
and ($\alpha_y, \delta_y$) correspond to the same celestial object.
Again, the identification threshold $d_2$ may depend
on various factors and should be estimated experimentally
for each catalogue studied.

\subsectionb{2.3}{Refining the WCT source catalogues}

To identify the cases described in \S\,2.1 and \S\,2.2, we applied
the above methods to three principal catalogues of visual binaries
-- WDS, CCDM, and TDSC -- varying the $d_1$ and $d_2$ threshold
values. We manually checked our preliminary results against the
Binary Star Database (BDB, Kaygorodov et al. 2012), which allows us
to visualize the catalogued data and estimated values of $d_1$ and
$d_2$ for those two tasks for every catalogue. After that, we used
our tools to compile the final lists of errors in the catalogues.

Note that the WDS (unlike the other two catalogues) does not provide
the coordinates for the additional component. It means that the task
(i), i.e. the detection of fictitious data, can be performed only
for those members of multiple (triple and higher multiplicity)
systems, which appear as the reference component in one pair and an
additional component in another one. For example, the WDS provides
the data for the following pairs of the system WDS~00013+6021: AB,
AC, AD, BD, and the task can therefore be performed only for
component B.
Also, the WDS provides two sets of $\theta, \rho$ values, for the
first and the last observations, and we used the latter one.


\begin{figure}[!tH]
\vbox{
\centerline{\psfig{figure=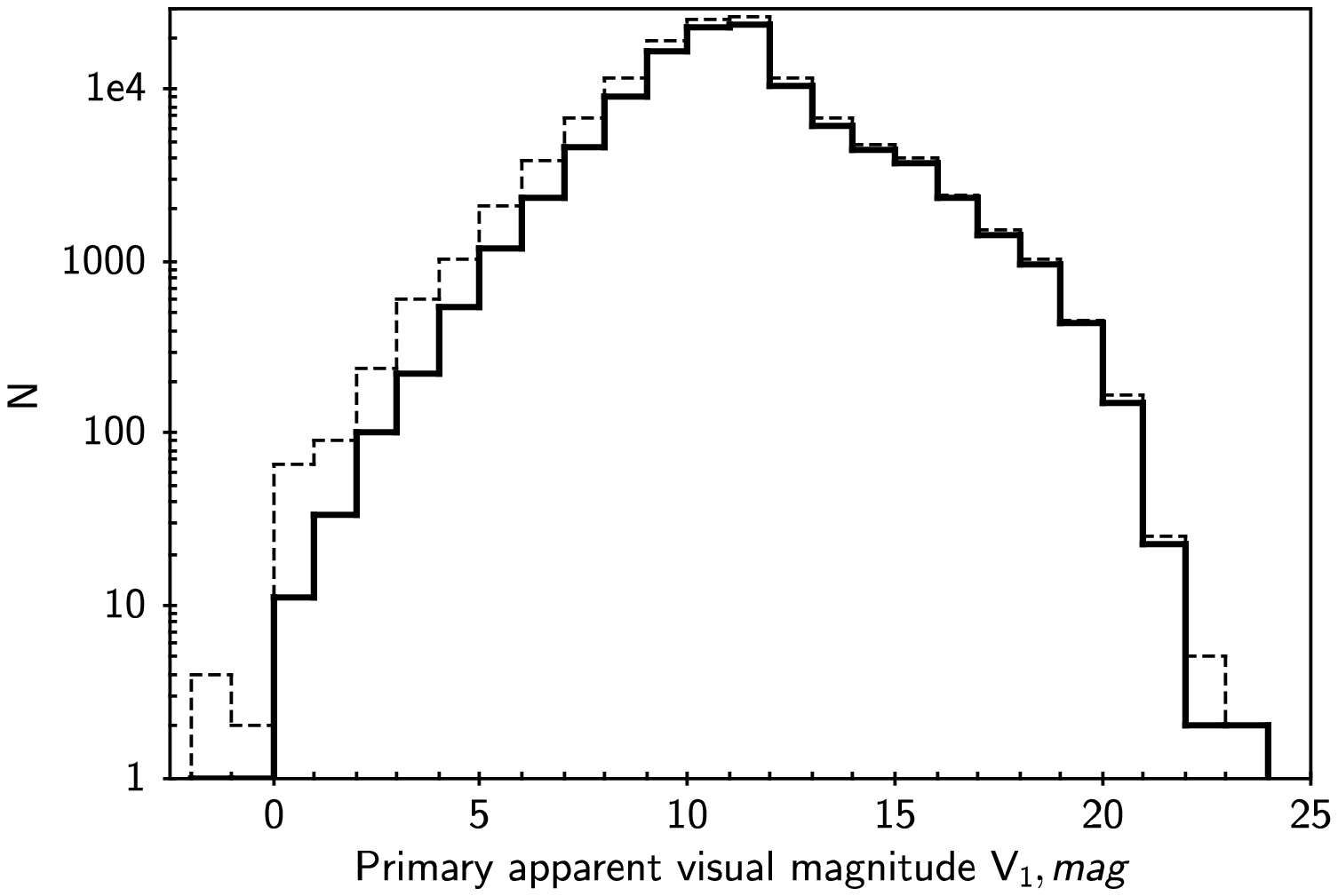,width=88mm,angle=0,clip=}}}
\vspace{1mm} \captionb{2} {Distribution of primary component
magnitudes of all WCT pairs (the dashed line) compared to that after
correction for multiples with only one brightest pair left for every
multiple system (the solid line).}
\end{figure}

We found $d_1$=8~arcsec for the WDS, with about 340 pairs in the
catalogue having inconsistent positional data for additional
components. The reasons for this inconsistency can be large orbital
motion in relatively close pairs (especially in the cases of large
difference of observing epochs), large proper motion difference in
optical pairs, confusion of two closely spaced objects, 180-degree
ambiguity in positional angle $\theta$, large and, consequently,
imprecisely defined separation $\rho$, and typos in the catalogues.

Estimating the $d_2$ value is a more difficult task, as it depends
on the surface density of stars in a given area. For the WDS, we
found $d_2 \approx 35$~arcsec. About 330 couples of objects have
smaller $d$. Most of these couples in fact represent a single
object, however, our analysis shows that some of them, especially
those with $d>15$~arcsec, can indeed represent different objects.

The CCDM catalogue, albeit smaller than the WDS, is based on a
similar set of observational data. On the other hand, the WDS is
constantly updated, whereas the CCDM contains observational data as
collected by 2002. That is why setting $d_1$ for the CCDM equal to
8~arcsec results in about 1000 pairs having the listed coordinates
of the secondary inconsistent with those based on the coordinates of
the primary, position angle, and separation. We estimate $d_2$ for
the CCDM to be of about 11~arcsec, and in 19 cases components are
included in two different binary/multiple systems in the catalogue.

The TDSC is a homogeneous catalogue, and it contains observations
performed with the same instrument. As a result, we found $d_1$ to
be about 1.2 arcsec for most of the objects. Only eight very wide
pairs are beyond this limit with the component separation $\rho$
exceeding 10~arcmin, and for the extreme case $d=27.4$~arcsec
(TDCS~56356 = WDS~20452-3120AB, $\rho$\,$=$\,$78$~arcmin). Because
of its homogeneity, no duplicate entries could be found in the TDSC,
at least at the level of $d_2 \approx 50$~arcsec. The only exception
is TDSC 29583~A = TDSC 29584~A.

Note also that our analysis revealed about 70 other errors (typos)
in the CCDM, and the list of the errors was submitted to VizieR.
Some errors were also found in the WDS, and appropriate reports were
sent to the authors of this permanently updated catalogue.


\begin{figure}[!tH]
\vbox{
\centerline{\psfig{figure=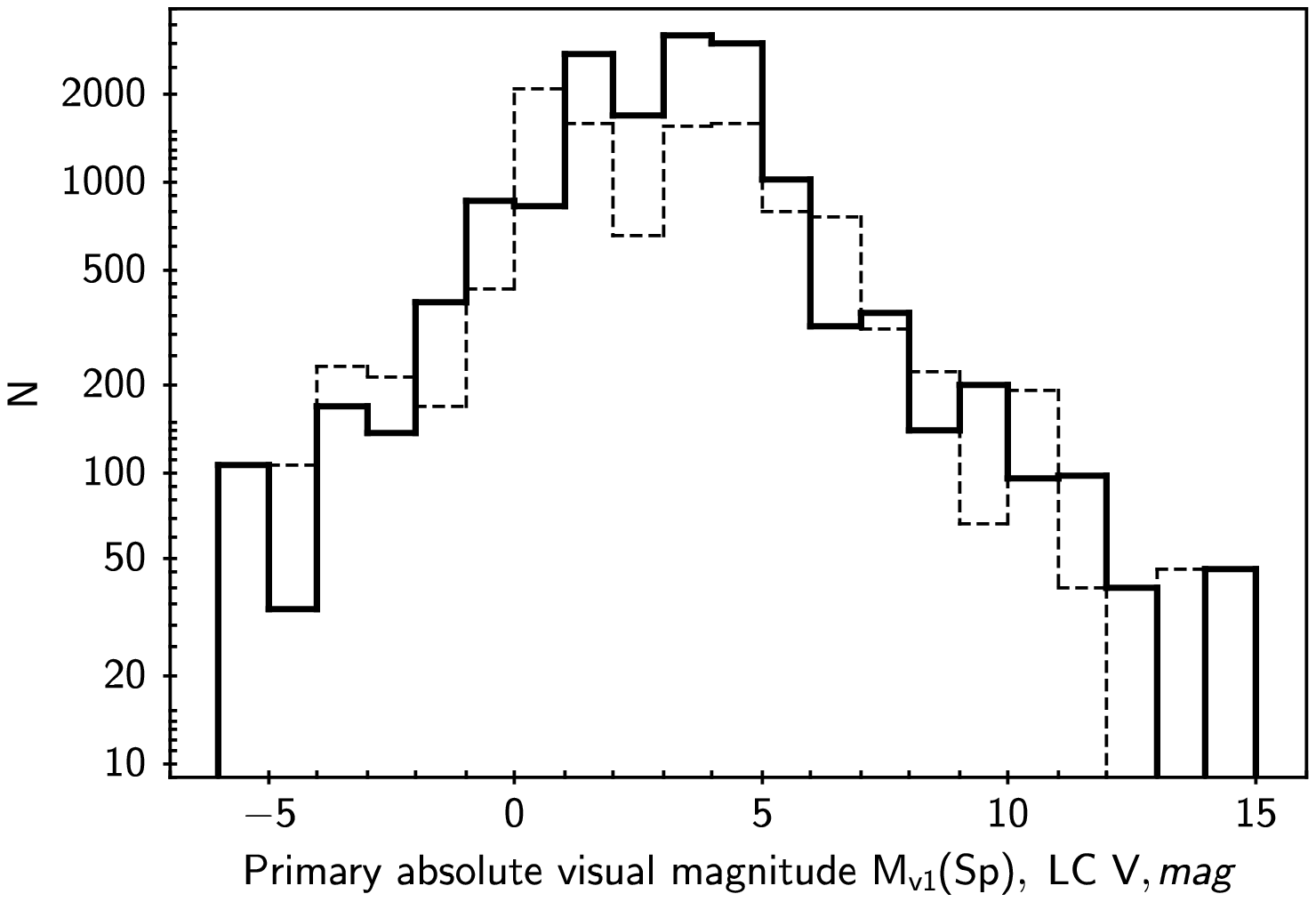,width=88mm,angle=0,clip=}}}
\vspace{1mm} \captionb{3} {Distribution of absolute visual
magnitudes of luminosity class V primaries. The dotted and solid
lines show the distributions of absolute magnitudes determined using
the calibrations of Strai{\v z}ys (1982) and those of Mamajek (2014)
and Pecaut \& Mamajek (2013), respectively. The visible depression
between magnitudes 2 and 3 is due to non-uniformity of spectral
classification.}
\end{figure}

\sectionb{3}{SELECTING DATA SAMPLE FOR STATISTICAL INVESTIGATION}

After taking into account errors in the WCT source catalogues as
described in the previous section, one needs to perform several more
steps to obtain  a data sample suitable for statistical
investigation.

\subsectionb{3.1}{Rounded up data removal}

The angular separations for the WCT pairs (see Fig.~1) are biased by
because they are rounded up. In the WDS, there is a certain number
of binaries detected interferometrically and not resolved by visual
methods. For these pairs, $\rho$ is set to be $-1.0$. Both in the
WDS and CCDM, the a$\rho$ is given with an accuracy of 1 dex. Hence
all stars with $\rho$ values less than approximately $0.05$~arcsec
are assigned zero separation (0.0), whereas the actual separations
of these pairs may  strongly differ. The same is true of the
separations between approximately 0.05 and 0.15~arcsec, which are
all listed as 0.1~arcsec in the catalogue.  Similarly, in the WDS
catalogue, all pairs separated by 1000~arcsec and more have been
assigned separations equal to 999.9~arcsec. In the histogram of the
component angular separation a prominent peak at 1.0~arcsec is
present, which is evidently of a similar nature. A separate analysis
of the $\rho$-distributions for the three source catalogues helps to
check these effects and remove the data that produce false peaks. At
this stage, slightly more than 126\,000 pairs remain in the sample.

\subsectionb{3.2}{Treatment of multiple systems}

More than 8200 systems in the WCT have multiplicity of 3 and more.
The pairs of these systems should be treated correctly because
otherwise stars of multiple systems (including the brightest ones)
would be included into the sample several times. This is illustrated
by Fig.~2. For every multiple system, we retain a single pair (the
brightest one) for further investigation. After this stage, still
more than 112\,000 pairs remain in the sample.

\subsectionb{3.3}{Removal of optical pairs}

The catalogues of visual binaries unavoidably contain some fraction
of non-physical pairs. These data should also be removed from our
sample. Some 4500 pairs are marked in the notes to the WDS catalogue
as non-physical. Moreover, we apply a "$1\%$" statistical filter
(Poveda et al. 1982) to remove systems that do not satisfy the
condition
\begin{equation}
\pi d^2 N(m_{v,2})<0.01,
\label{equ:filter}
\end{equation}
where $d$ is the angular separation between the components with
apparent magnitudes $m_{v,1}$ and $m_{v,2}$; $N(m_{v,2})$ is the
number of stars brighter than $m_{v,2}$ per unit area in the
direction of the primary with galactic coordinates $(l,b)$ (adopted
after Allen 1977). We can thus expect the systems retained in our
sample to have the probability of being random close projected pairs
of less than $1\%$.

\subsectionb{3.4}{Luminosity classes of the sample stars}

Deriving the distributions of some of the parameters of visual
binaries involves the use of photometric properties of the stars. To
this end, it is desirable to consider a star sample uniform with
respect to luminosity class. Spectral classifications are available
for more than 63\,000 pairs of the WCT, and luminosity classes are
assigned to less than a half of the stars. Among stars with known
luminosity classes, about 18\,000 pairs have two dwarf components.
Note that the spectra of both components are rarely available and
are used mainly to exclude from further consideration the pairs with
degenerate and other peculiar components. Basically, we treated the
total spectrum of a pair as that of the primary.

However, the fraction of erroneous spectral classifications is known
to be rather significant. According to estimates by Mironov (2015,
private communication), up to $20 \%$ of  classifications of stars
of luminosity class V and III in the Hipparcos catalogue may be
wrong. We can check this for the WCT stars with trigonometric
parallaxes by comparing the absolute visual magnitudes determined
for the primaries of these stars using (i) visual magnitude,
parallax and extinction, and (ii) spectral classification. Fig.~3
compares the distributions of absolute visual magnitudes of
luminosity-class V primaries for binaries of our sample, determined
using the calibrations proposed by Strai{\v z}ys (1982) and by
Mamajek (2014) and Pecaut \& Mamajek (2013). The two distributions
appear to basically agree with each other.

\enlargethispage{-4mm} There are more than 4000 pairs with both
spectral classification (assuming luminosity class V) and
trigonometric parallaxes available. We estimated interstellar
extinction $A_v$ for these stars using the cosecant law (Parenago
1940),

\begin{equation}
A_v(r,b) = \frac{a_0 \beta}{|\sin b|}
\left[1 - e^{-\frac{r |\sin b|}{\beta}}\right],
\label{equ:extinction}
\end{equation}
where  $r$ is the heliocentric distance; $b$, the galactic latitude;
$\beta$, the scale height, and $a_0$, the extinction per kpc in the
direction of an object located in the Galactic plane behind the
absorbing layer. Here we adopt $a_0$~=~1.6 mag\,kpc$^{-1}$ and
$\beta$\,=\,114~pc Sharov (1963). Malkov \& Kilpio (2002) have shown
that this law, although rather old, represents interstellar
extinction quite adequately within relatively close vicinity of the
Sun ($r\aplt1$kpc), where most of the known orbital binaries reside.


\begin{figure}[!tH]
\vbox{
\centerline{\psfig{figure=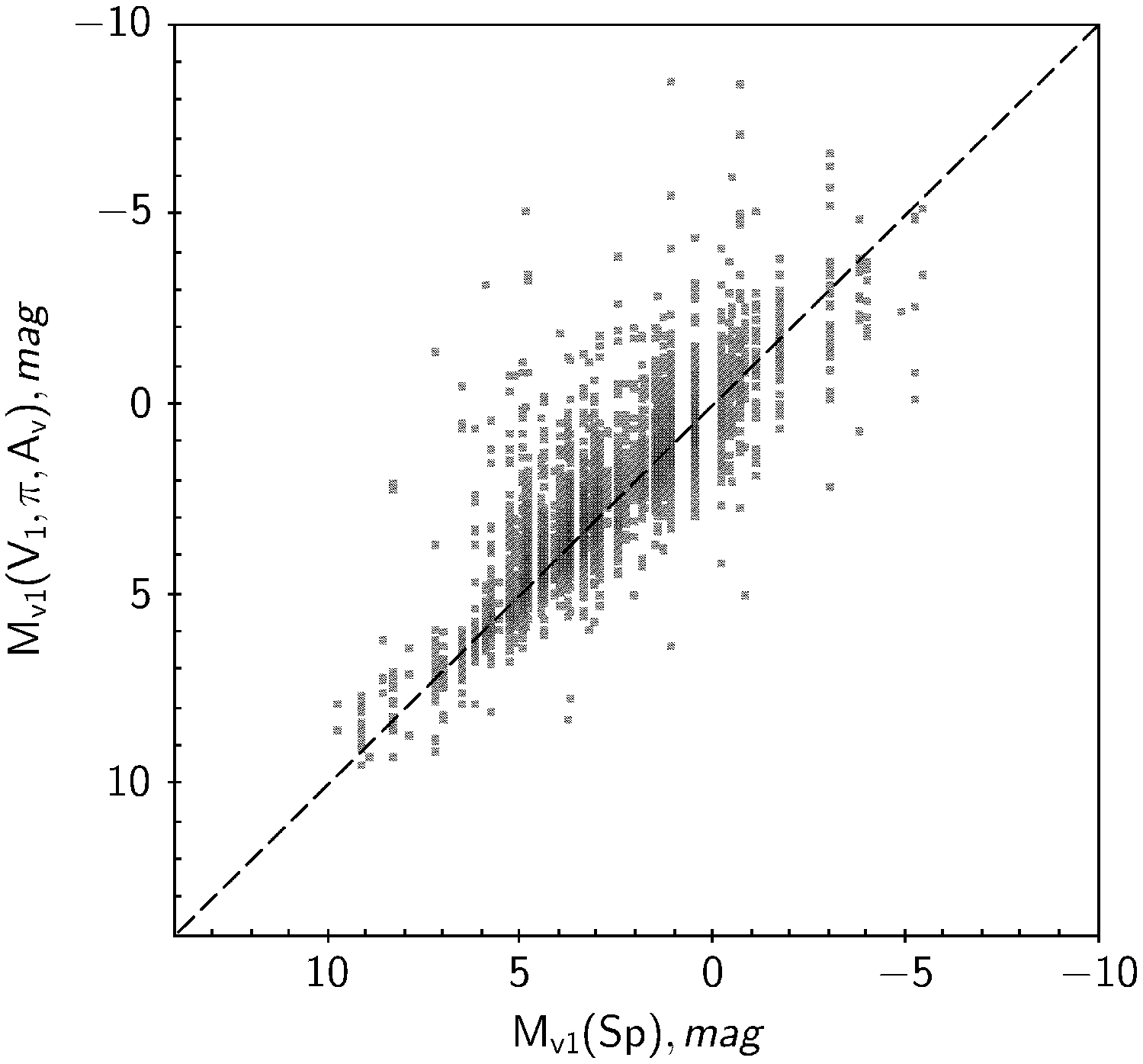,width=100mm,angle=0,clip=}}}
\vspace{1mm} \captionb{4} {Luminosity class V pairs: the
spectroscopic absolute magnitudes of primaries versus the absolute
magnitudes determined from trigonometric parallax, apparent
magnitude, and interstellar extinction. Primaries with $\Delta
M_v>2$\,mag are excluded from further investigation as possibly
misclassified luminous stars. }
\end{figure}

Fig.~4 compares the primary absolute magnitudes $M_v$ determined
using the two methods described above. We find the standard
deviation of the $\Delta M_v$ distribution to be $1.3$\,mag. Deb \&
Chakraborty (2014) estimated the intrinsic scatter of the $\Delta
M_v$ differences obtained in the process of spectral
re-classification of stars within 100 pc to be $2.0$\,mag. If the
brighter star is misclassified as a dwarf, the absolute magnitude
determined from the apparent magnitude and parallax should exceed
the value inferred from the spectral type: $\Delta M_v=M_v(m_v,\pi,
A_v)-M_v(Sp)>0$. We exclude the objects with $\Delta M>2.0$\,mag
from further consideration as possibly misclassified luminous stars.
A part of them (more than $40\%$) have large fractional parallax
errors (greater than $50\%$, and amounting to hundreds of percent in
some of the cases), which can also be a source of discrepancy.
However, other stars have reasonably good parallaxes and rather poor
spectrum quality in SIMBAD (typically, D in the scale where A is the
best quality and E is the worst). Some of them are close binaries;
few, perhaps, are of peculiar nature. We plan to check these stars
manually in the course of further investigation. Eliminating objects
with possibly erroneous luminosity classification decreases the size
of our sample by almost 400 stars.

\subsectionb{3.5} {Restrictions due to selection effects}
At this point, we have a sample of main-sequence pairs with known
spectral classification and parallaxes that has been cleaned of
supposedly erroneous data. It contains 3676 pairs, and hereafter we
refer to this dataset as ``refined''. The sample is inevitably
distorted by selection effects of various nature. These effects can
be subdivided into ``observational'' (those depending on observing
techniques) and ``evolutionary'' (those due to stellar evolution
processes). However, these two groups of selection effects are not
independent because many observational parameters are affected by
evolution. In this part of our study we consider the most obvious
``observational'' selection effects.


\begin{figure}[!tH]
\vbox{
\centerline{\psfig{figure=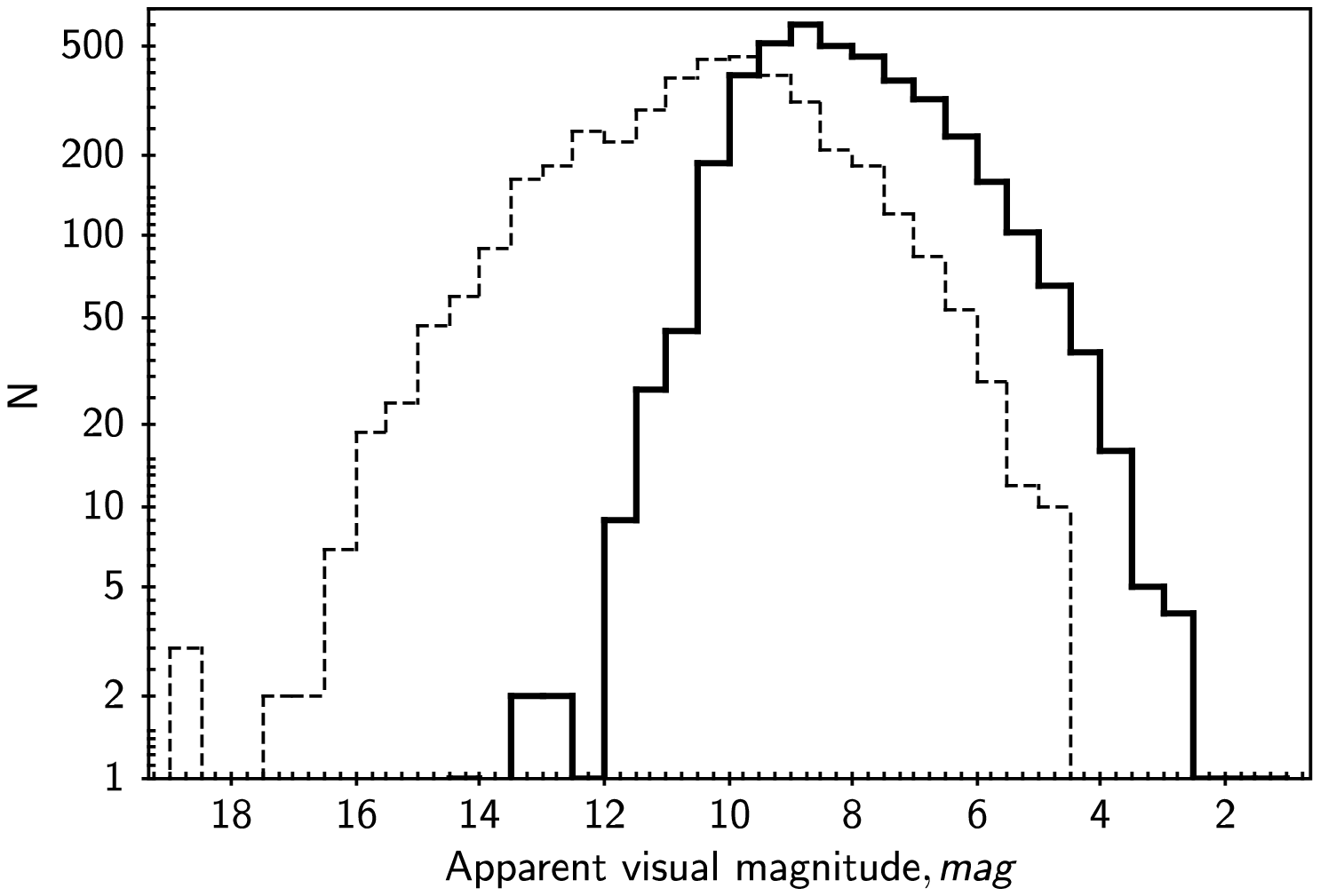,width=88mm,angle=0,clip=}}}
\vspace{1mm} \captionb{5} {Distributions of apparent magnitudes for
the primaries (the solid line) and secondaries (the dashed line) in
the refined set of 3676 pairs. The effect of selection by secondary
magnitude is clearly seen: the number of systems starts to decrease
at $10.5$\,mag.}
\end{figure}

We restrict our dataset by the secondary magnitude of $10.5$\,mag
(Fig.~5) as the number of pairs with the fainter secondaries
decreases.


\begin{figure}
\includegraphics[width=6.0cm]{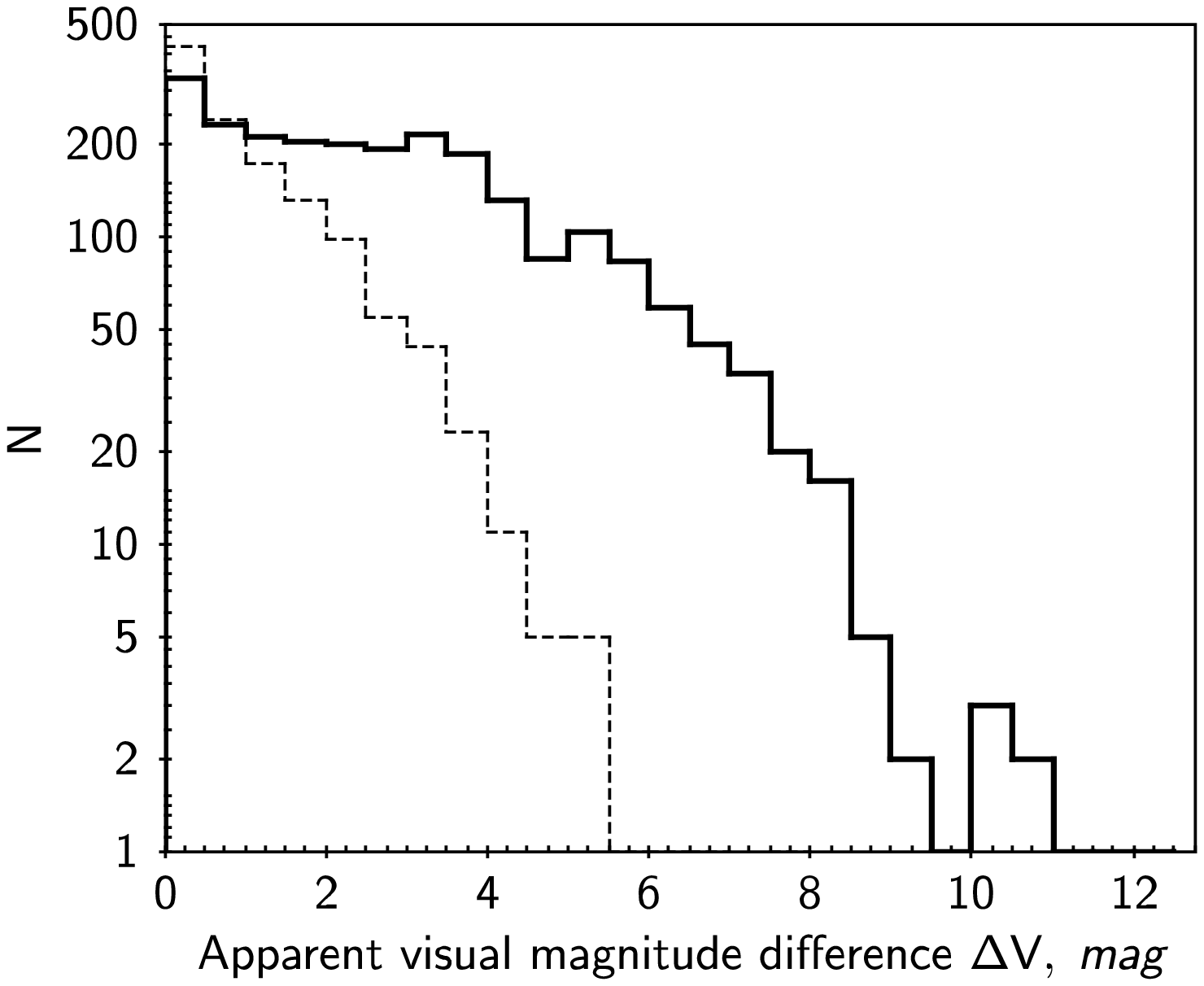}
\includegraphics[width=6.0cm]{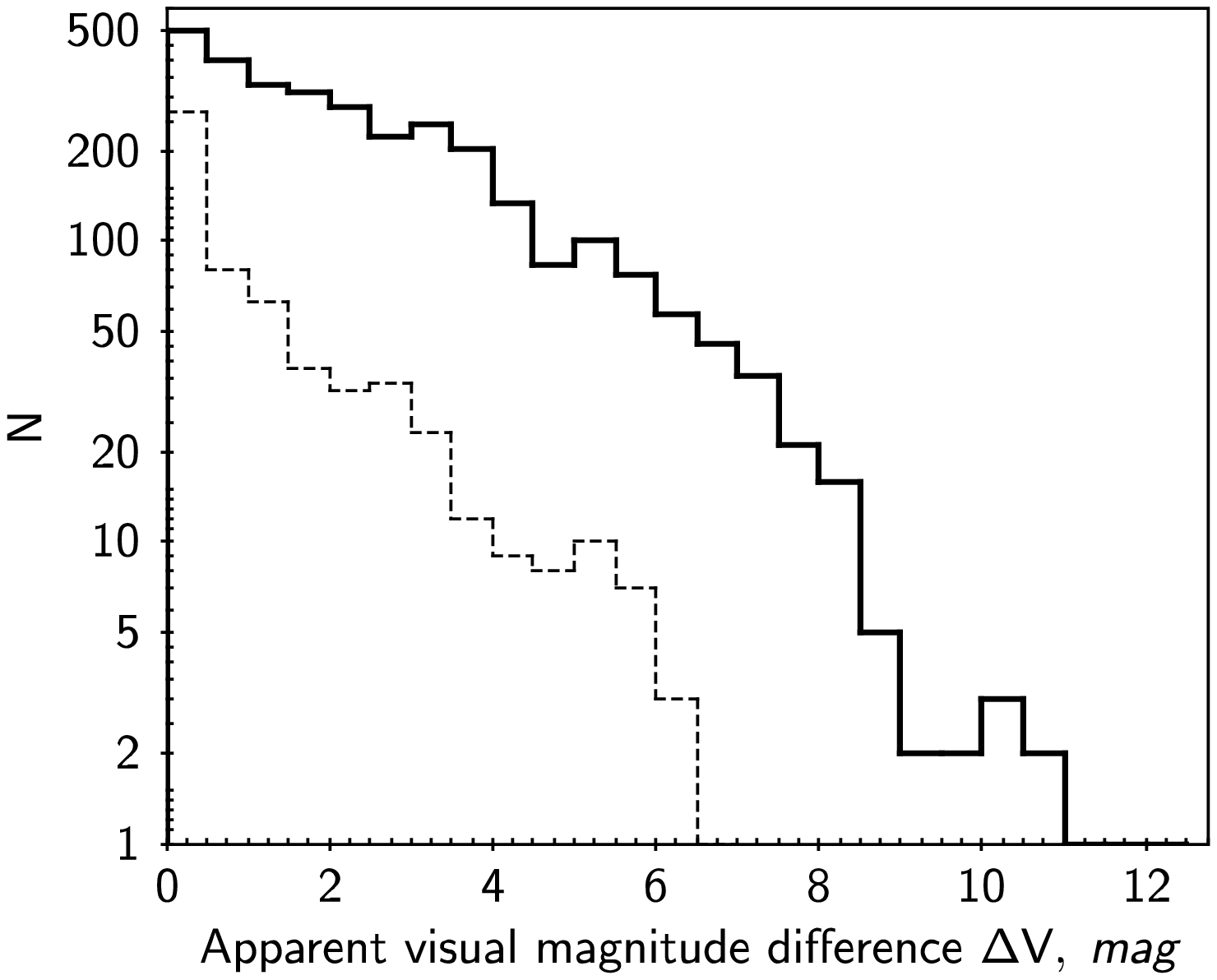}
\vspace{5pt} \captionb{6}{ Distributions of the difference in
magnitudes of the components, $\Delta V$, in the refined set of 3676
pairs. The left-hand panel shows the distribution for pairs with
$\rho > 1$ arcsec (the solid line; the distribution is almost flat
(no dependence) at $\Delta V<4.0$\,mag) and for those with $\rho <
1$ arcsec (the dashed line; the dependence is strong for all $\Delta
V$). The right-hand panel shows the distribution for stars with
$V_1<9.5$\,mag (the solid line; the distribution is almost flat (no
dependence) at $\Delta V<4.0$\,mag) and for those with $V_1 >
9.5$\,mag (the dashed line, there is no unbiased $\Delta V$ domain.)
}
\end{figure}

The chance for a star to be detected as a visual binary depends on
the angular separation between the components, primary magnitude,
and magnitude difference. Let us consider how the discovery of a
pair with a certain magnitude difference depends on $\rho$ (the
left-hand panel in Fig.~6) and on $V_1$ (the right-hand panel in
Fig.~6). It was established that in both cases the dataset is
divided in two parts rather strictly. For $\rho > 1$~arcsec, there
is almost no correlation between $\rho$ and $\Delta V$ up to $\Delta
V \simeq 4$\,mag (except for a small peak close to 0); similarly,
for $V_1 < 9.5$\,mag, the dependence is weak up to $\Delta V \simeq
4$\,mag. Otherwise, for smaller $\rho$ and fainter $V_1$, the
correlation between these parameters and $\Delta V$ is strong and
obvious and there is no $\Delta V$ range without such a correlation.

This was the reason to further limit our dataset to the systems with
$\rho > 1$ arcsec, $V_1 <9.5$\,mag, and $\Delta V < 4$\,mag, in
order
to avoid the domains of the space of parameters where the
sample is obviously incomplete. The resulting dataset is much
smaller than the previous (refined) one, and contains 998 pairs.
Hereafter we refer to this sample as ``restricted''.


\begin{figure}[!tH]
\vbox{
\centerline{\psfig{figure=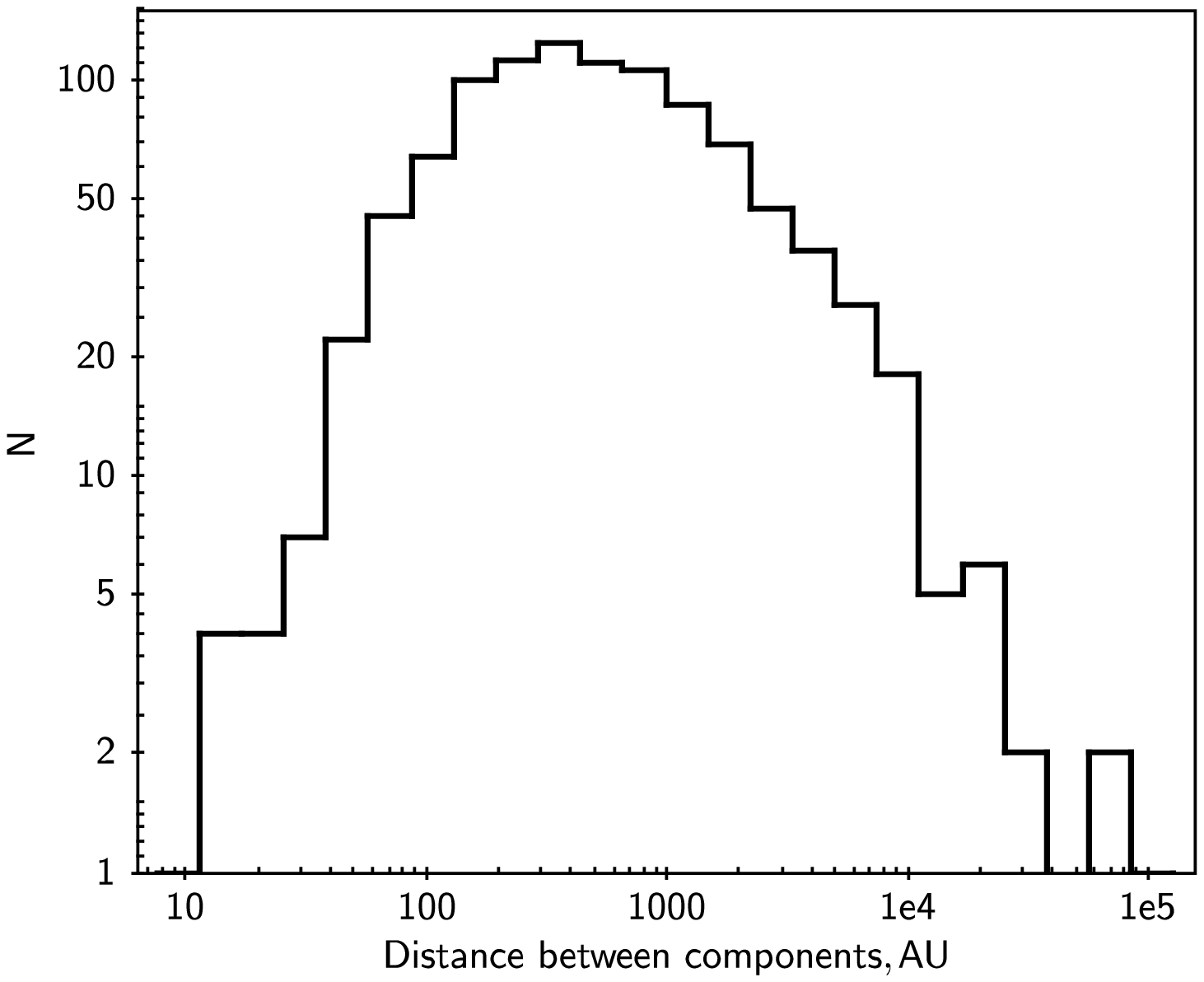,width=88mm,angle=0,clip=}}}
\vspace{1mm} \captionb{7} {Distribution of the distance between the
components for the restricted sample of 998 visual binaries. The
dataset is still not free of selection effects and cannot be viewed
as representative of the initial or present-day distribution of
binary parameters.}
\end{figure}

We do not consider here the selection effects involving orbit
orientation or spatial location (because of the implicit uniformity
of the distribution of the corresponding quantities for field
stars). We plan to incorporate selection effects due to stellar
evolution at the next stage of our investigation (see below).

\sectionb{4}{RESULTS AND PROSPECTS}


\begin{figure}[!tH]
\vbox{
\centerline{\psfig{figure=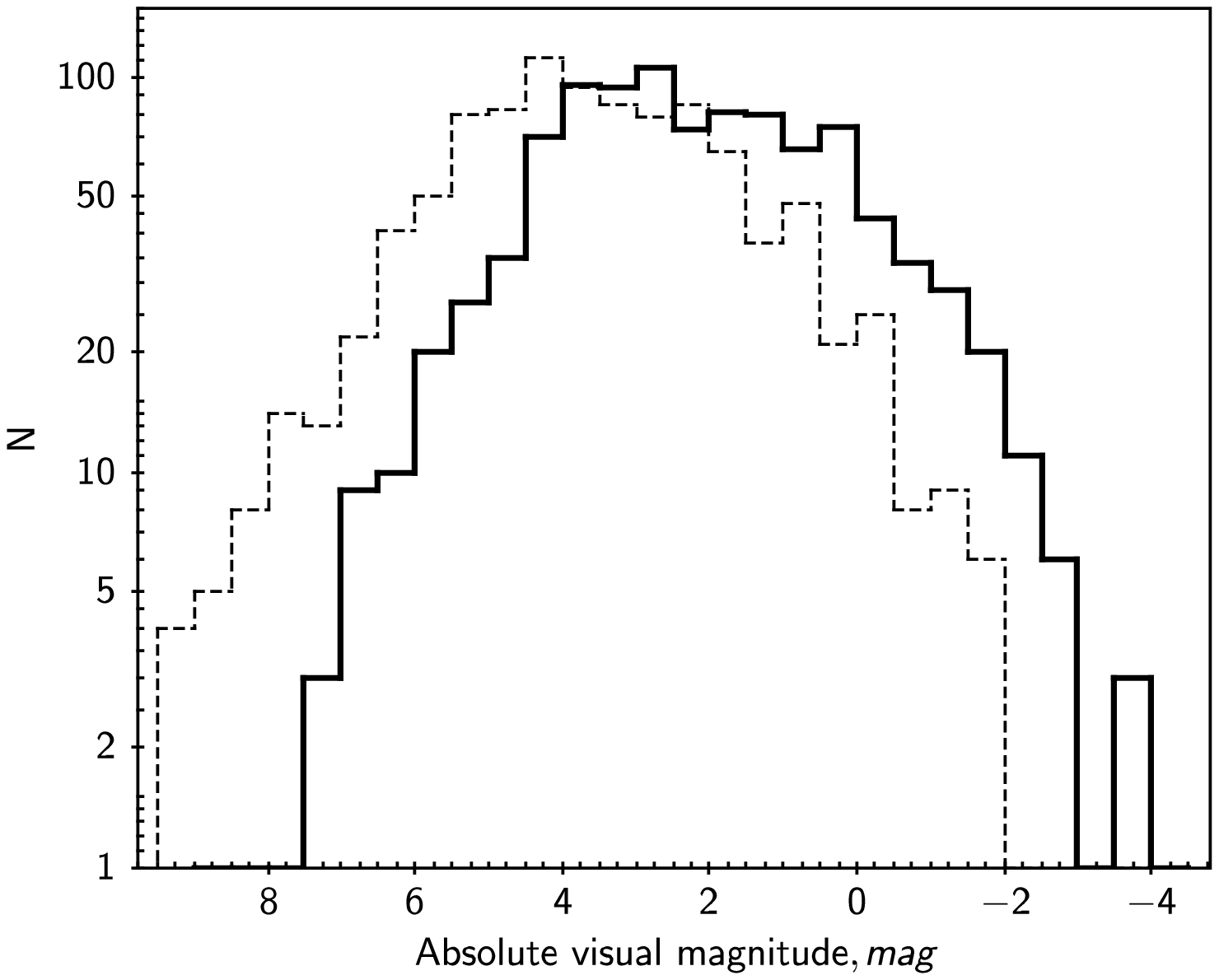,width=88mm,angle=0,clip=}}}
\vspace{1mm} \captionb{8} {Distributions of the absolute magnitudes
for the primaries (the solid line) and secondaries (the dotted line)
in the restricted sample of 998 visual binaries. The dataset is
still not free of selection effects and cannot be viewed as
representative of the initial or present-day distribution of binary
parameters. }
\end{figure}

In the process of this  study we identified and corrected a number
of errors in the comprehensive dataset of visual binary stars, the
WCT, originating from its source catalogues, as well as the errors
due to several other factors causing dataset distortion. We obtained
a number of samples of visual binaries characterized by different
data amount and quality. The refined set of data contains 3676
(presumably) physical visual pairs of luminosity class V with known
angular separations, magnitudes of the components, spectral classes,
and parallaxes, and is hopefully free from technical biases. We also
obtained a restricted sample of 998 pairs free from certain obvious
observational biases (those due to the probability of binary
discovery related to secondary magnitude, angular separation,
primary magnitude, and magnitude difference).

Known parallaxes allow one to easily convert angular separations
into actual distances between the components (Fig.~7). It would be
more difficult, however, to convert the distribution of distances
into that of semimajor axes given the lack of an established model
of eccentricity distribution of binaries. There are few analytical
approaches, like the thermal distribution $f(e)\sim2e$ (first
proposed by Ambartsumian 1937) which is expected if the orbits were
distributed in phase space according to a function of energy
exclusively, or $f(e)=$\,const. Some dynamical simulations lead to
more complicated formulae (see, for instance, Bate 2009 or
Stamatellos \& Whitworth 2009). Different observational datasets
seem to follow different distributions (Dupuy \& Liu 2011; Shatsky
2001; Tokovinin 1998; etc.). With a particular model adopted one may
pass to the distribution of semimajor axes using statistical factors
calculated assuming that the orbital inclinations of binaries are
distributed uniformly. For instance, such a factor is equal to 1.26
for the thermal distribution of eccentricities, and 0.98 for the
hypothesis of circular orbits (all $e=0$) (Kouwenhoven et al. 2008).

Fig.~8 shows the distribution of the absolute magnitudes of the
primary and secondary stars in the restricted sample.

Note that all of the samples discussed here, including the
restricted one, are distorted by the selection effects and cannot be
considered representative of the initial or present-day distribution
of binary parameters. In further investigations, we plan to generate
simulated samples based on various initial distributions of masses
and orbital elements, and incorporating the effects of stellar
evolution and observational selection. We will improve the data
quality and, hopefully, increase the number of objects in certain
datasets by thoroughly investigating the spectral types and
observational background of selected stars. Our final aim is to find
the distributions over physical parameters, such as $M_1$,
$M_2/M_1$, and $a$, for nascent binaries.

\sectionb{5}{CONCLUSIONS}

We analyzed the most comprehensive dataset of visual binary stars,
the WCT, for the purpose to make it useful for the statistical
investigation of wide binaries. We eliminated certain technical
factors causing the dataset distortion and obtained a number of
samples of visual binaries, characterized by different data amount
and quality. The refined set of data contains 3676 (presumably)
physical visual pairs of luminosity class V with known angular
separations, magnitudes of the components, spectral types, and
parallaxes, and is hopefully free from technical biases. To avoid
the domains of parameter space where the whole sample is obviously
incomplete, we also obtained a restricted sample of 998 pairs (1)
having known two-dimensional spectral classification and parallaxes,
(2) free from certain obvious observational biases due to the
probability of binary discovery, (3) with separations between the
components $\rho > 1$ arcsec, (4) with apparent visual magnitudes
$V_1 <9.5$~mag, $V_2 <10.5$~mag, and (5) with differences in
magnitudes of the components $\Delta V < 4$~mag. For this dataset,
the distributions of the separations between the components and the
absolute magnitudes are known. In our further investigations we plan
to analyze the distributions of initial/present-day parameters for
wide binaries.

\thanks{We thank Alexei Mironov and Pavel Kaygorodov for useful suggestions and discussion.
This work was supported in part by the Russian Foundation for Basic
Research (grant No.~15-02-04053) and the Presidium of the Russian
Academy of Sciences (Program P-41). This research has made use of
the VizieR catalogue access tool and the SIMBAD database operated at
CDS, Strasbourg, France, the Washington Double Star Catalog
maintained at the U.S. Naval Observatory, and NASA's Astrophysics
Data System Bibliographic Services. \enlargethispage{-3mm}}

\References

\refb Allen K. W. 1977, \textit{Astrophysical Quantities}
(Translated from the 3rd revised and suppl. English edition), Mir,
Moskva

\refb Ambartsumian V. A. 1937, Astron. Zhurn., 14, 207

\refb Bate M. R. 2009, MNRAS, 392, 590

\refb Deb D., Chakraborty P. 2014, PASA, 31, 46

\refb Dommanget J., Nys O. 2002, VizieR On-line Data Catalog: I/274

\refb Dupuy T. J., Liu M. C. 2011, ApJ, 733, 122

\refb Fabricius C., H{\o}g E., Makarov V. et al. 2002, A\&A, 384,
180

\refb Isaeva A. A., Kovaleva D. A., Malkov O. Yu. 2015, Baltic
Astronomy, 24, 157

\refb Kaygorodov P., Debray B., Kolesnikov N., Kovaleva D., Malkov
O. 2012, Baltic Astronomy, 21, 309

\refb Kouwenhoven M. B. N., Brown A. G. A., Goodwin S. P., Portegies
Zwart S. F., Kaper L. 2008, Astron. Nachr., 329, 984

\refb Malkov O. Yu., Kilpio E. Yu. 2002, Ap\&SS, 280, 115

\refb Mamajek E. 2014,\\
\url{www.pas.rochester.edu/~emamajek/EEM_dwarf_UBVIJHK_
colors_Teff.dat}

\refb Mason B. D., Wycoff G. L., Hartkopf W. I., Douglass G. G.,
Worley C. E. 2014, VizieR On-line Data Catalog: B/wds

\refb Parenago P. P. 1940, Bull. Sternberg Astron. Inst., 4

\refb Pecaut M. J., Mamajek E. E. 2013, ApJS, 208, 9

\refb Poveda A., Allen C., Parrao L. 1982, ApJ, 258, 589

\refb Sharov A. S. 1963, Astron. Zh., 40, 900

\refb Shatsky N. I. 2001, A\&A, 380, 238

\refb Stamatellos D., Whitworth A. P. 2009, MNRAS, 392, 413

\refb Strai\v{z}ys V. 1982, {\it Metal-Deficient Stars}, Mokslas
Publishers, Vilnius

\refb Tokovinin A. A. 1998, Astron. Lett., 24, 178

\refb Vereshchagin S., Tutukov A., Iungelson L., Kraicheva Z.,
Popova E. 1988, Ap\&SS, 142, 245

\end{document}